# Strong PUFs from arrays of resonant tunnelling diodes

B. Astbury, I.E. Bagci, T. McGrath, J. Sexton, M. Missous, U. Roedig, R. Bernardo-Gavito, and R.J. Young

In this work, we design and implement a strong physical uncloneable function from an array of individual resonant tunnelling diodes that were previously described to have a unique response when challenged. The system demonstrates the exponential scalability of its responses when compared to the number of devices present in the system, with an expected large set of responses while retaining a 1:1 relationship with challenges. Using a relatively small set of 16 devices, 256 responses are shown to have promising levels of distinctness and repeatability through multiple measurements.

*Introduction:* With the advent of interconnected societies in a modern world with a large-scale Internet of Things (IoT), our reliance on software, hardware and networks has increased dramatically. Globally, the dependence on the security of these systems further increases with our adoption of these networks, where the market is expected to continue to expand rapidly in the coming years [1]. The trust we place in the security of these systems may be unfounded, with reports of attacks using hundreds of thousands of compromised IoT devices [2]. With an increase in the number of attacks comes an expected increase in success.

One solution to address the security and authentication of users is to exploit physically uniqueness from a system in the growing field of physical uncloneable functions (PUFs). PUFs, as opposed to digital keys, boast uniqueness that is uncloneable. Various implementations exist ranging from optical to electronic-based devices. These can be categorised by their inherent security and implementation method as being either weak or strong [3].

One type of PUF proposed recently uses a quantum effect in resonant tunnelling diodes (RTDs) to create a single unique response from uncontrollable atom-scale differences in semiconductor manufacture. The quantum-confinement PUF (QC-PUF) [4] has a single challenge-response pair (CRP) per device, derived from a peak in its current-voltage trace (see Fig. 1a). While this weak PUF can be used to generate keys, its limited number of CRPs leaves it exposed to attackers and thus requires protected access (a protected housing, for example). For a lightweight authentication system, a larger set of CRPs, which increase exponentially with some system parameter, is required to ensure each response, when used, is not compromised.

The solution proposed herein consists of combining RTDs in arrays such that each set of RTDs can output a much larger CRP set.

*Theory:* RTDs provide a unique fingerprint from naturally-occurring atom-scale variations in the semiconductor manufacturing process.

RTDs were previously used by Roberts et al. as a weak PUF [4]. They are a physical manifestation of a finite quantum well (QW) with potential barriers in a conducting channel. The devices are created by sandwiching a thin, narrow band-gap semiconductor between two wide band-gap semiconductors to act as the main QW structure. Beyond the tunnelling barriers lies highly-doped narrow band-gap regions of semiconductor which form the electron source and sink [5].

These devices operate under two electron transport regimes, depending on the voltage that is applied across them. The transition between these regimes provides their characteristic electronic behaviour, which is illustrated in Fig. 1a. At low voltages, quantum resonant tunnelling through the classically forbidden region dominates, for which the device finds its name. The conduction level rises as a single confined energy level comes into resonance with electrons that have the required energy to pass through [6]. A shift in transport mechanism happens when the energy level passes out of resonance, resulting in a shift to classical conduction. This is known as thermionic emission, where electrons with the most thermal energy pass over the top of the tunnelling barrier [6].

The transition between these regimes is a unique response in the current-voltage characteristic of every RTD, which is defined by the first allowed energy level within the QW structure. This is highly sensitive to small variations, such as well width. The single peak that results from the shift in conduction mechanism (Fig. 1) serves as the unique characteristic that is different for each device.

To create a large database of responses, characterisation of how two devices interact in a single circuit must show that the properties of one device acts on that of the second device. The impedance of devices, for example, results in a combined shift in peak from that of a single device. It is seen that different devices cause different shifts due to slight variations in resistances at different points in a single current-voltage response of the devices. The result is that each combination of devices is unique to the devices within them, and the base RTDs are unique due to variations in manufacture.

*Methodology*: Networks of diodes were connected with switches, allowing the current path to be controlled. This is illustrated in Fig 1c. It allows the uniqueness of the convoluted response of multiple devices to be measured, to ascertain how the CRPs for the system grows with the number of elements within it. Rectangular arrays of devices were built, with $n$ columns containing $m$ RTDs. The aim is to increase the number of responses by more than the simple product, $m \times n$, to $m^n$. For an array with $3 \times 4$ elements, for example, this would provide 81 permutations (i.e. chains of 4 RTDs in series), with each current-voltage characteristic containing 4 peaks in each permutation (see Fig. 1b).

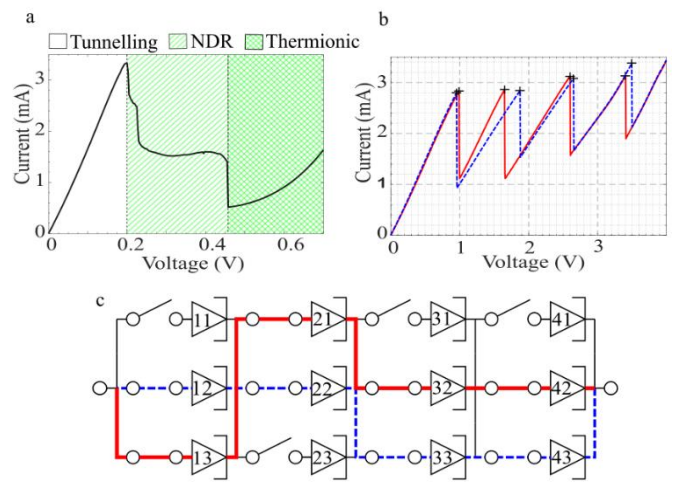

**Fig. 1** *Current-voltage characteristics of single and arrayed RTDs.*
*a* - A trace showing the current-voltage curve for a single RTD, sectioned to show the prominent transport mechanisms and the negative differential resistance (NDR). *b* - Current-voltage curves for two different linear chains of 4 RTD devices connected in series. *c* - A 3x4 array of individually switched RTDs. The circuit representation shows two possible current paths (blue or red) through the system, for different switch configurations.

The total number of permutations of current paths through RTDs linked in series through switched arrays with different array dimensions is illustrated in Fig. 2. The parameters of the array can be tuned to maximise the outputs and the points by which authentication can be made (peak positions). As can be seen in Fig. 2, the number of permutations increases much faster as the number of devices per column increase. This corresponds to having more devices to swap and more peaks per permutation. Due to the exponential nature of the increase, the optimal value of rows is the natural logarithm, $e$. In terms of a physical number of devices per column, this corresponds to 2 or 3 rows. For example, 100 devices can be configured in a $10 \times 10$ array, giving $1 \times 10^{10}$ responses, or a $2 \times 50$ array, giving far more ($1.125 \times 10^{15}$) responses. Optimising the array's configuration makes a significant difference to the total number of responses therefore, increasing the time in which it takes an adversary to characterise all the responses of a single system.

A 4 x 4 array of RTDs was constructed where each diode is in series with a switch. Each RTD was individually wire-bonded using 15 µm gold wire to the circuit, with 16 switches between the diodes controlled by an Arduino Due. RTD characteristics were measured using a Keithley 2602B SMU. Permutations were run consecutively with multiple-subsequent measurements of each to test similarity.

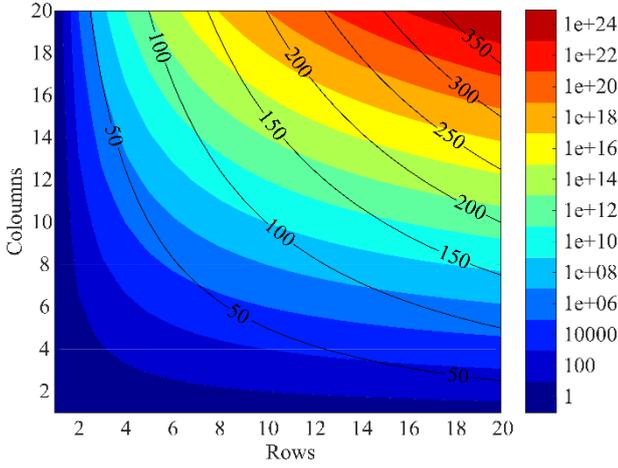

**Fig. 2** *The number of permutations of an array (colour scale) as a function the rows and columns it contains. The black contours show the total number of RTDs in the array.*

*Results*: The axes of the current-voltage curve were divided into equidistant bins with a permutation's response characterised by the bin/s containing peak/s. This system of bins was used to determine distinctness and similarity of the permutations denoted by uniqueness and robustness. Uniqueness is defined as how distinct each permutation is from one another and uses the expected output of a permutation as its comparison point.

A 1 is given if a compared peak is found in the same bin and 0 if different, the term is denoted by equation (1). The uniqueness is given by equation (2) as a sum of difference metrics and divided by the number of summed terms. Robustness is given as the complement of metric given from (2) using repeat measurements as $R_i, R_j$. Both metrics are created as probabilities that the next measurement will conform to the expected output.

$$diff(R_i, R_j) = \begin{cases} 0, & R_i = R_j \\ 1, & R_i \neq R_j \end{cases} \quad (1)$$

$$\frac{1}{k}\sum_{i,j=1, i \neq j}^{k} diff(R_i, R_j) \quad (2)$$

Table 1 gives results for a 4x4 array with 256 permutations having 256 bins. For each peak position, on average a high percentage, unique measurement is seen with 3 out of 4 peaks registering greater than 90% uniqueness. This is coupled with a similar robustness for each of the peak positions of ~80%. Peak 1 was found to have the lowest uniqueness, due to the near-linear dependence between current and voltage in the resonant tunnelling regime.

Comparing each 4-peak permutation to a subsequent permutation, if one or more peaks are distinct then the permutation is distinct. This gives a uniqueness measure that convolutes all four peak positions. The same process can be repeated for subsequent measurements of the same permutation to ascertain the robustness of 4-peak measurements. As expected, an increase in overall uniqueness and a decrease in robustness is observed in Table 1. This finding was expected, as convoluting peak positions provide more points for a distinct signature, but measurement variations also increase the likelihood of failure of measuring true-positive results.

**Table 1:** The result of uniqueness and robustness measurements performed on a 4 x 4 array with 256 permutations and 4 peaks. Results are first shown for each of the four individual peaks, then their average and finally for using all peaks simultaneously.

| Array Size | Bins | Identifier | Uniqueness (%) | Robustness (%) |
|---|---|---|---|---|
| 4x4 | 256 | Peak 1 | 76.8 | 83.9 ± 12.6 |
| | | Peak 2 | 95.0 | 78.4 ± 10.4 |
| | | Peak 3 | 94.4 | 82.8 ± 10.9 |
| | | Peak 4 | 92.3 | 81.0 ± 14.8 |
| | | Average | 85.6 | 81.5 ± 12.3 |
| | | **4-Peak** | **99.7** | **60.7 ± 17.2** |

*Discussion:* Overall, the system uniqueness is high, with a 4-peak value over 99%. Real implementations of a system like this in cryptographic applications would require higher values that could be achieved by scaling the array size and increasing the number of bins used. The relatively low robustness of 60% would necessitate measuring the response to challenges multiple times to reduce false-negative rates. Modifying the system implementation to reduce noise in the measurement process would inherently increase the robustness of the measurements, and also allow the bin size to be decreased, in turn increasing the uniqueness of the responses. The implications of results provided herein are evocative of a system useful in providing increased security to IoT devices with low power, cost-effective results.

*Conclusion:* We have demonstrated an exponentially growing design for a strong PUF system using resonant tunnelling diodes. The number and form of the responses can be tailored to the security requirement of an application by maximising the unique outputs of the system. This system shows promising results in its ability to unique, repeatable responses in scalable devices. The system is comprised of nanoscale elements that can be implemented in CMOS technology [7], with a relatively small gate count for the entire device. We believe the small scale, low power and cost requirements make the implementation of this technology well suited to IoT devices and other applications of small embedded systems.

*Acknowledgements:* RJY acknowledges support by the Royal Society through a University Research Fellowship (UF110555 and UF160721). This material is based upon work supported by the Air Force Office of Scientific Research under award number FA9550-16-1-0276. This work was also supported by grants from The Engineering and Physical Sciences Research Council in the UK (EP/K50421X/1 and EP/L01548X/1) and the Royal Society through a Brian Mercer award.

[Copyright statement goes here
Submitted:
doi:
One or more of the Figures in this Letter are available in colour online]

B. J. Astbury, R. Bernardo-Gavito, T. McGrath and R. J. Young (*Physics Department, Lancaster University, Lancaster, United Kingdom, LA1 4YB*)

E-mail: r.j.young@lancaster.ac.uk or r.bernardogavito1@lancaster.ac.uk

I.E. Bagci, U. Roedig (*School of Computing and Communications, InfoLab21, Lancaster University, Lancaster, United Kingdom, LA14WA*)

J. Sexton, M. Missous, (*School of Electrical and Electronic Engineering, The University of Manchester, Manchester, United Kingdom M13 9PL*)

**References**

1 Lund, D., Morales, M.: 'Worldwide and Regional Internet of Things (IoT) 2014 – 2020 Forecast : A Virtuous Circle of Proven Value and Demand', IDC Anal. Futur., 2014, (May), p. 29.

2 Antonakakis, M., April, T., Bailey, M., et al.: 'Understanding the Mirai Botnet', Proc. 26th USENIX Secur. Symp., 2017, pp. 1093–1110.

3 Gao, Y., Ranasinghe, D.C., Al-Sarawi, S.F., Kavehei, O., Abbott, D.: 'Emerging Physical Unclonable Functions with Nanotechnology', IEEE Access, 2016, 4, pp. 61–80.

4 Roberts, J., Bagci, I.E., Zawawi, M.A.M., et al.: 'Using Quantum Confinement to Uniquely Identify Devices', Sci. Rep., 2015, 5, (1), p. 16456.

5 Zawawi, M.A.M., Ian, K.W., Sexton, J., Missous, M.: 'Fabrication of submicrometer INGAAS/ALAS resonant tunneling diode using a trilayer soft reflow technique with excellent scalability', IEEE Trans. Electron Devices, 2014, 61, (7), pp. 2338–2342.

6 Ricco, B., Azbel, M.Y: 'Physics of resonant tunneling. The one-dimensional double-barrier case', Phys. Rev. B, 1984, 29, (4), p. 1970-1981.

7 Lavieville, R., Triozon, F., et al: 'Quantum dot made in metal oxide silicon-nanowire field effect transistor working at room tempterature', Nano. Lett., 2015, 15, (5), p. 2958-2964.